\documentstyle[12pt]{article}
\input epsf.sty
\def\ZZZ{{\hbox{ Z\kern-1.6mm Z}}}

\newcommand{\OO}{{\cal O}}

\newcommand{\LL}{{\cal L}}

\newcommand{\wt}{\widetilde}
\newcommand{\wh}{\widehat}

\newcommand{\SSS}{{\cal S}}

\newcommand{\be}{\begin{equation}}
\newcommand{\ee}{\end{equation}}
\newcommand{\ben}{\begin{eqnarray}\displaystyle}
\newcommand{\een}{\end{eqnarray}}
\newcommand{\refb}[1]{(\ref{#1})}
\newcommand{\p}{\partial}
\newcommand{\sectiono}[1]{\section{#1}\setcounter{equation}{0}}

\def\one{{\hbox{ 1\kern-.8mm l}}}
\def\zero{{\hbox{ 0\kern-1.5mm 0}}}

\begin{document}
{}~
{}~
\hfill\vbox{\hbox{hep-th/0505122}
}\break

\vskip .6cm
\begin{center}
{\Large \bf
Stretching the Horizon of a Higher Dimensional Small Black Hole
}

\end{center}

\vskip .6cm
\medskip

\vspace*{4.0ex}

\centerline{\large \rm
Ashoke Sen}

\vspace*{4.0ex}

\centerline{\large \it Harish-Chandra Research Institute}

\centerline{\large \it  Chhatnag Road, Jhusi,
Allahabad 211019, INDIA}

\centerline{E-mail: ashoke.sen@cern.ch,
sen@mri.ernet.in}

\vspace*{5.0ex}

\centerline{\bf Abstract} \bigskip

There is a general scaling argument that shows that the entropy of a small
black hole, representing a half-BPS excitation of an elementary heterotic
string in any dimension, agrees with the statistical entropy up to an
overall numerical factor. We propose that for suitable choice of field
variables the near horizon geometry of the black hole in
$D$ space-time dimensions
takes the form of
$AdS_2\times S^{D-2}$ and demonstrate how this ansatz can be used to calculate
the numerical factor in the expression for the black hole entropy if we know
the higher derivative corrections to the action. We illustrate this by
computing the entropy of these black holes in a theory where we modify the
supergravity action by adding the Gauss-Bonnet term.
The black hole entropy computed this way is finite and has the right
dependence on the charges in accordance with the general scaling argument, but
the overall numerical factor
does not agree with that computed from the statistical
entropy except for $D=4$ and $D=5$.
This is not surprising in view of the fact that we do not
use a fully supersymmetric action in our analysis; however this analysis
demonstrates that higher derivative corrections are capable
of stretching the horizon
of a small black hole in arbitrary dimensions.

\vfill \eject

\baselineskip=18pt

\tableofcontents

\sectiono{Introduction and Summary} \label{sintro}

For heterotic string compactification to space-time dimension $D\le 9$
with at least $N=2$ supersymmetry, we have a set of half-BPS states in
the spectrum of elementary string states\cite{rdabh0,rdabh1}. The simplest
examples of such states are elementary string states wound
$w$ times on a circle
and carrying $n$ units momentum along the same circle. For a fixed set of
charge quantum numbers the degeneracy of these BPS states grow as
$\exp(4\pi\sqrt{nw})$. Thus we could assign a statistical entropy
$4\pi\sqrt{nw}$ to these states.

Given this result it is natural to ask if the same answer can be reproduced
by computing the entropy of a black hole carrying the same charge quantum
numbers\cite{thooft,9309145,9401070,9405117}.
The supergravity solution representing this state has
vanishing area of the event horizon and hence vanishing entropy. However
the curvature and other field strengths grow as we approach the horizon,
and hence we expect higher derivative ($\alpha'$)
corrections to modify the geometry near the
horizon. In contrast the string coupling is small near the horizon and hence
we can ignore string loop corretions to leading order.
Using the symmetries of the tree
level effective action of string theory
it was shown, first in \cite{9504147} for four dimensional black
holes and then in \cite{9506200} for higher dimensional black holes,
that for these
black holes the $\alpha'$ corrections produce an entropy
of the form $a\sqrt{nw}$ where $a$ is a purely numerical
constant.\footnote{Some aspects of higher derivative corrections to these
solutions have been discussed in \cite{9408040,9509050}.}
This
has the same dependence on $n$ and $w$ as the statistical entropy, and is
independent of the asymptotic values of various moduli
fields as is the case
for the statistical entropy. Computation of the numerical coefficient
$a$ however requires a detailed knowledge of the higher
derivative corrections to the effective action, and was not carried
out in these papers.

Recently there has been renewed interest in the
problem\cite{0409148,0410076,0411255,0411272,0501014,0502126,0502157,0504005}
due to the
observation of
Dabholkar
that upon inclusion of a special class of
higher derivative terms in the effective action of four dimensional
heterotic string theory, -- obtained by supersymmetrizing the curvature
squared
term in the action, --
the near horizon geometry of the solution
becomes $AdS_2\times S^2$. This gives a finite entropy, and in fact
we get exactly the right factor of $4\pi$
for the coefficient $a$ appearing in the expression
for the black hole entropy. This is in precise agreement with the
statistical entropy. The developments which made this computation possible
were a series of papers\cite{9602060,9603191,
9801081,9812082,9904005,9906094,9910179,0007195,0009234,0012232}
which supersymmetrized the curvature squared
term in the effective action, and used it to compute corrections to the
black hole solution / entropy carrying both electric and magnetic charges.
The case
of elementary string states, which carry only electric charge, then
follows from the results of \cite{9602060,9603191,9801081,
9812082,9904005,9906094,9910179,0007195,0009234,0012232}
by setting the magnetic charge to zero.
It is however not clear why the other higher derivative corrections which
have not been included in the analysis do not
affect the result. The general expectation is that there is some kind
of non-renormalization theorem that prevents further correction to the
entropy formula for the black hole, although no such theorem has been proven
to this date.

Given the success of this program for heterotic string compactification to four
dimensions, it is natural to ask whether we can extend this analysis
to higher dimensions.
Here we are at a disadvantage since the construction of a fully
supersymmetric version of the action containing curvature squared terms, along
the line of \cite{9602060,9603191,9801081,
9812082,9904005,9906094,9910179,0007195,0009234,0012232},
has not been done. Nevertheless one might wonder if it
is possible to use the intuition gained from the four dimensional problem
to guess the general structure of the horizon of these higher
dimensional black holes. This is the problem we undertake in this paper.
Our ansatz for the near horizon geometry is $AdS_2\times
S^{D-2}$. This background is characterized by several parameters, --
the radii of $AdS_2$ and $S^{D-2}$,
the constant values of the scalar fields near the horizon, and the
flux of the gauge fields through $AdS_2$. We then demonstrate how, given
a classical effective action, we can determine these parameters by solving
a set of algebraic equations. We illustrate this procedure by computing
all the parameters in a theory where we add to the usual supergravity
action the Gauss-Bonnet term\cite{rzwiebach,9610237}.
We get a finite area event horizon
and finite entropy of the black hole, although the overall numerical
factor in the expression for the black hole
entropy does not agree with that in the expression for the
statistical entropy except in four and five dimensions.
Given that we have
not even included terms which are related to the Gauss-Bonnet term via
supersymmetry transformation, the result is hardly surprising. However the
analysis demonstrates that the higher derivative terms do have the capability
of modifying the near horizon geometry of these black holes to produce
a finite answer for the entropy.

The rest of the paper is organised as follows.
We work in units $\hbar=c=1$ and $\alpha'=16$.
In section \ref{sreview}
we review the arguments of \cite{9504147,9506200,9712150} which establish
the correct dependence of the entropy of the black hole on the charge
quantum
numbers $n$ and $w$ characterizing the black hole.
In the process of doing so we also restate the arguments of \cite{9506200}
along the lines of \cite{9504147,0411255} that makes explicit the
general validity of the argument.
In section
\ref{scomplete} we propose an ansatz for the modification of the near
horizon geometry of the black hole due to $\alpha'$ corrections to the
effective action and study its effect on the black hole entropy. In
particular we show how our ansatz reduces the problem of finding the
near horizon geometry of the black
hole solution in the higher derivative theory
to solving a set of algebraic equations, and the computation
of the entropy associated with the solution to evaluation of an
algebraic expression. We also demonstrate this procedure explicitly
in the context of an action where we modify the supergravity action by
addition of the
Gauss-Bonnet term. This gives rise to a regular near horizon solution and
finite entropy of the black hole although the numerical coefficient
computed from this action
does not agree with the one computed from statistical entropy for
$D\ne 4,5$.
Finally
in section \ref{sinter} we discuss some aspects of the construction
of the full solution that
interpolates between the near horizon geometry and the geometry at large
radial distance.

\sectiono{Supergravity Solution for Two Charge Black Holes
and its Near
Horizon Limit} \label{sreview}

In this section we shall review the black hole
solutions of the supergravity equations of motion representing
elementary string states in toroidally compactified heterotic string
theory, its near horizon limit, and the scaling argument leading to the
general form of the entropy associated with these black
holes\cite{9504147,9506200}.

In
order to keep our discussion simple, we
shall here consider only a special class of black hole solutions
representing a heterotic string wound on a circle. We
work with
heterotic string theory compactified on $T^n\times S^1$,
$T^n$ being an arbitrary $n$-torus and $S^1$ being a circle of
coordinate radius
$\sqrt{\alpha'}=4$. Let us denote by $x^M$ ($0\le M\le 9$) the set of all
coordinates, by
$x^\mu$ ($0\le\mu\le D-1$, $D=9-n$) the coordinates along the
non-compact directions, by
$x^m$ ($D\le m\le 8$) the coordinates of $T^n$ and by $x^9$ the
coordinate
along $S^1$. We also denote by $G^{(10)}_{MN}$,
$B^{(10)}_{MN}$ and $\Phi^{(10)}$ the ten dimensional string
metric, anti-symmetric tensor field and dilaton
respectively.
For
the description of the black hole solution under study we shall only need
to consider non-trivial configurations of the fields $G^{(10)}_{\mu\nu}$,
$B^{(10)}_{\mu\nu}$,
$G^{(10)}_{9\mu}$, $G^{(10)}_{99}$, $B^{(10)}_{9\mu}$ and $\Phi^{(10)}$.
We freeze all other field components to trivial background values, and
define:\footnote{Our convention for normalization
of the dilaton is the
same as that
in \cite{9504147,9506200}, \i.e. $e^{\Phi}$ represents the square
of the effective
closed
string coupling constant.}
 \ben \label{e3}
 && \Phi = \Phi^{(10)} - {1\over 2} \, \ln (G^{(10)}_{99})\, , \qquad
S=e^{-\Phi}\, , \qquad
 T = \sqrt{G^{(10)}_{99}}\, , \nonumber \\
&& G_{\mu\nu} = G^{(10)}_{\mu\nu} - (G^{(10)}_{99})^{-1} \,
G^{(10)}_{9\mu}
\, G^{(10)}_{9\nu}\, ,
\nonumber \\
&& A^{(1)}_\mu = {1\over 2} (G^{(10)}_{99})^{-1} \, G^{(10)}_{9\mu}\, ,
\qquad
A^{(2)}_\mu = {1\over 2} B^{(10)}_{9\mu}\, , \nonumber \\
&& B_{\mu\nu} = B^{(10)}_{\mu\nu} - 2(A^{(1)}_\mu A^{(2)}_\nu -
A^{(1)}_\nu A^{(2)}_\mu)
\, .
 \een
The low energy effective action involving these fields is then given
by
\ben \label{e4-}
\SSS &=& {1\over 32\pi} \int d^D x \, \sqrt{-\det G} \,
S \, \bigg[ R_G
+ S^{-2}\, G^{\mu\nu} \, \p_\mu S \p_\nu S -  T^{-2}
\, G^{\mu\nu} \, \p_\mu T \p_\nu T
\nonumber \\
&&- {1\over 12}  G^{\mu\mu'}
G^{\nu\nu'} G^{\rho\rho'} H_{\mu\nu\rho} H_{\mu'\nu'\rho'} - T^2 \,
G^{\mu\nu} \, G^{\mu'\nu'} \, F^{(1)}_{\mu\mu'}
F^{(1)}_{\nu\nu'} - T^{-2} \,
G^{\mu\nu} \, G^{\mu'\nu'} \, F^{(2)}_{\mu\mu'}
F^{(2)}_{\nu\nu'}\bigg] \, , \nonumber \\
\een
where $R_G$ is the scalar curvature computed from the string metric
$G_{\mu\nu}$ and
\ben \label{e4a}
&& F^{(a)}_{\mu\nu} = \p_\mu A^{(a)}_\nu - \p_\nu A^{(a)}_\mu\, , \quad
a=1,2\, , \nonumber \\
&& H_{\mu\nu\rho} = \left[ \p_\mu B_{\nu\rho} + 2 \left( A_\mu^{(1)}
F^{(2)}_{\nu\rho} - A_\mu^{(2)}
F^{(1)}_{\nu\rho}\right) \right] + \hbox{cyclic permutations of $\mu$,
$\nu$, $\rho$}\, .
\een
The overall normalization constant of $1/32\pi$ is completely arbitrary
and could be absorbed into a redefinition of $S$ by an multiplicative constant.
$T$ has been normalized so that the $T$-duality transformation
takes the form:
\be \label{e4c}
T\to {1\over T}\, , \qquad F^{(1)}_{\mu\nu}\to F^{(2)}_{\mu\nu}, \qquad
F^{(2)}_{\mu\nu}\to F^{(1)}_{\mu\nu}\, .
\ee

We shall focus on field configurations for which
\be \label{ehmn}
H_{\mu\nu\rho} = 0\, .
\ee
For studying these configurations we can use the restricted action
\ben \label{e4+}
\SSS &=& {1\over 32\pi} \int d^D x \, \sqrt{-\det G} \,
S \, \bigg[ R_G
+ S^{-2}\, G^{\mu\nu} \, \p_\mu S \p_\nu S -  T^{-2}
\, G^{\mu\nu} \, \p_\mu T \p_\nu T
\nonumber \\
&& - T^2 \,
G^{\mu\nu} \, G^{\mu'\nu'} \, F^{(1)}_{\mu\mu'}
F^{(1)}_{\nu\nu'} - T^{-2} \,
G^{\mu\nu} \, G^{\mu'\nu'} \, F^{(2)}_{\mu\mu'}
F^{(2)}_{\nu\nu'}\bigg] \, ,
\een
obtained by setting $H_{\mu\nu\rho}=0$ in \refb{e4-}.

Eq.\refb{e4+} expresses the relevant part of the
action in terms of the string metric
$G_{\mu\nu}$. The same expression, written in terms of the canonical metric
\be \label{edefgmn}
g_{\mu\nu} = S^\gamma \, G_{\mu\nu}\, , \qquad \gamma={2\over D-2}\, ,
\ee
takes the form:
\ben \label{e4}
\SSS &=& {1\over 32\pi} \int d^D x \, \sqrt{-\det g} \, \bigg[ R_g - {1\over
(D-2)
} \, S^{-2}\,  g^{\mu\nu} \, \p_\mu S \p_\nu S -  T^{-2}\,
g^{\mu\nu} \, \p_\mu T \p_\nu T
\nonumber \\
&&   - S^\gamma T^2 \,
g^{\mu\nu} \, g^{\mu'\nu'} \, F^{(1)}_{\mu\mu'}
F^{(1)}_{\nu\nu'} - S^\gamma T^{-2} \,
g^{\mu\nu} \, g^{\mu'\nu'} \, F^{(2)}_{\mu\mu'}
F^{(2)}_{\nu\nu'}\bigg] \, . \nonumber \\
\een
With this choice of overall normalization convention the
Newton's constant is given by
\be \label{e5.0h}
G_N = 2\, .
\ee
Note that this action is invariant under
\be \label{einv1}
T\to e^\beta T\, , \quad S \to e^\lambda S, \, \quad
A^{(1)}_{\mu} \to e^{-\beta-\gamma\lambda/2} A^{(1)}_{\mu}, \quad
A^{(2)}_{\mu} \to e^{\beta-\gamma\lambda/2}
A^{(2)}_{\mu}\, , \quad
g_{\mu\nu}\to g_{\mu\nu}\, ,
\ee
for arbitrary real numbers $\beta$ and $\lambda$. As we shall argue later, the
transformation generated by $\beta$ is an exact symmetry of the {\it tree level}
string effective action.

We now consider an heterotic string wound $w$ times along
the circle $S^1$ labelled by $x^9$ and
carrying $n$ units of momentum along the same circle.
These solutions may be obtained by beginning with the D-dimensional
Schwarzschild solution and then by applying a solution generating
transformation\cite{9109038,9411187}.
The solutions relevant for us are
given by\cite{9506200}
\ben \label{e6-}
ds_{string}^2
&\equiv& G_{\mu\nu} dx^\mu dx^\nu = - (F(\rho))^{-1} \rho^{2\alpha}
dt^2 +
d\vec{x}^2 \, , \nonumber \\
&& \rho^2 \equiv \vec x^2\, , \qquad F(\rho)
\equiv(\rho^\alpha + 2W) (\rho^\alpha+2N)
\, , \qquad \alpha\equiv D-3 \, ,
\nonumber \\
S &=& (F(\rho))^{1/2} \, \rho^{-\alpha} \, ,
\nonumber \\
T &=& \sqrt{ (\rho^\alpha+2N) / (\rho^\alpha + 2W)} \, , \nonumber \\
A^{(1)}_{t}  &=& -{N \over (\rho^\alpha+2N)} + \hbox{constant} \, ,
\nonumber \\
A^{(2)}_{t} &=& -
{W\over (\rho^\alpha + 2W)} + \hbox{constant} \, ,
\een
where $N$ and $W$ are two arbitrary parameters labelling the solution.
The canonical metric associated with this solution is given by
\ben \label{e6++}
ds_c^2 &\equiv& g_{\mu\nu} dx^\mu dx^\nu = S^\gamma
ds_{string}^2\nonumber \\
&=& - (F(\rho))^{{\gamma\over 2}-1} \rho^{\alpha(2-\gamma)}
dt^2 +
(F(\rho))^{{\gamma\over 2}} \,
\rho^{-\gamma\alpha} d\vec{x}^2 \, .
\een

The
solution \refb{e6-}
has the property that asymptotically the fields $S$ and $T$ approach
1. We shall be interested in a more general class of solutions for which
asymptotically
\ben \label{e5}
&& g_{\mu\nu} \to \eta_{\mu\nu} \nonumber \\
&& S\to g^{-2}, \qquad T\to R/4\,  .
\een
Here
$g$ is the value of the $D$ dimensional string coupling and $R$ denotes
the asymptotic radius of $S^1$ measured in the string metric. Such a solution
can be obtained from \refb{e6-} by applying the transformation \refb{einv1}
with
\be \label{einv3}
e^\lambda=g^{-2}, \qquad e^\beta=R/4\, .
\ee
This gives the new solution
\ben \label{e6+}
ds_c^2
&=& - (F(\rho))^{{\gamma\over 2}-1} \rho^{\alpha(2-\gamma)} \,
dt^2 +
(F(\rho))^{{\gamma\over 2}} \,
\rho^{-\gamma\alpha} d\vec{x}^2 \, ,
\nonumber \\
&& \rho^2 \equiv \vec x^2\, , \qquad F(\rho)
\equiv(\rho^\alpha + 2W) (\rho^\alpha+2N)
\, , \qquad \alpha\equiv D-3 \, ,
\nonumber \\
S &=& g^{-2}\, (F(\rho))^{1/2} \, \rho^{-\alpha} \, ,  \nonumber \\
T &=& {R\over 4}\,
\sqrt{ (\rho^\alpha+2N) / (\rho^\alpha + 2W)} \, , \nonumber \\
A^{(1)}_{t}  &=& -{4\over R}\, g^\gamma\,
{N \over (\rho^\alpha+2N)} + \hbox{constant} \, ,
\nonumber \\
A^{(2)}_{t} &=& -{R\over 4}\, g^\gamma\,
{W\over (\rho^\alpha + 2W)} + \hbox{constant} \, .
\een
The string metric and the gauge field strengths
associated with this solution is given by
\ben \label{estring}
ds_{string}^2 &=& S^{-\gamma} ds_c^2 = - g^{2\gamma}\,
(F(\rho))^{-1} \rho^{2\alpha}
dt^2 +
g^{2\gamma} \, d\vec{x}^2 \, , \nonumber \\
F^{(1)}_{\rho t}  &=& {4N\over R}\, g^\gamma\,
{\alpha \rho^{\alpha-1} \over (\rho^\alpha+2N)^2} \, ,
\nonumber \\
F^{(2)}_{\rho t} &=& {RW\over 4}\, g^\gamma\,
{\alpha \rho^{\alpha-1}\over (\rho^\alpha + 2W)^2} \, .
\een

{}From the definitions given in \refb{e3} it
follows that the gauge fields
$A^{(1)}_\mu$ and $A^{(2)}_\mu$ have as their sources
respectively the momentum and winding charge along $S^1$.
Since the asymptotic field configurations $F_{\mu\nu}^{(1)}$
and $F_{\mu\nu}^{(2)}$ are
proportional to $N$ and $W$ respectively, $N$ must be proportional to
the momentum quantum number $n$ and $W$ must be proportional to the
winding number $w$. We can find the constants of proportionality by
comparing the mass of the black hole with that of the elementary string
state carrying the same charges. For this we note that the black hole mass
$M$
can be read out by using the following asymptotic form of the $tt$
component of the canonical metric:
\be \label{egtt}
g_{tt} \simeq - 1 + {16\pi G_N \over (D-2) \Omega_{D-2}}
{M\over \rho^\alpha}
\, ,
\ee
where
\be \label{edefom}
\Omega_{D-2}\equiv 2 \, \pi^{(D-1)/2} \left/ \Gamma\left({D-1\over 2}
\right) \right.
\ee
is the volume of the unit $(D-2)$ sphere and
$G_N$ is the Newton's constant. Using \refb{e5.0h} we see that the mass
associated with the solution \refb{e6+} is given by:
\be \label{emass}
M = {(D-2) (2-\gamma) \, \Omega_{D-2} (N+W) \over 32\pi}
= {(D-3) \, \, \Omega_{D-2} (N+W) \over 16\pi} .
\ee
On the other hand, the mass of the BPS state of
an elementary string carrying $n$ units
of momentum and $w$ units of winding charge along $S^1$, measured in the
canonical metric, is given by
\be \label{emass2}
M = g^\gamma \left( {n\over R} + {w \, R \over 16} \right) \, .
\ee
Since $N$ and $W$ are expected to be proportional to $n$ and $w$
respectively, comparison of \refb{emass} and \refb{emass2} yields
\be \label{enwcomp}
N = {16\pi \over (D-3) \, \Omega_{D-2}} \, g^\gamma\, {n\over R}\, ,
\qquad W = {\pi \over (D-3) \, \Omega_{D-2}} \, g^\gamma\, w \, R\, .
\ee

The (singular) horizon of the solution \refb{e6+} is located at $\rho=0$.
It is easy to see that the area of the horizon, measured in the
canonical metric $g_{\mu\nu}=S^\gamma \, G_{\mu\nu}$, goes to zero
as $\rho\to 0$. Thus the black hole entropy also vanishes in this
approximation.

The question that we would like to address is: do higher
derivative ($\alpha'$) corrections
or string loop corrections modify this result? The effect of string loop
corrections is easy to estimate. {}From \refb{e6+} we see that the
square of the effective
string coupling, -- $S^{-1} = g^2 \rho^\alpha /\sqrt{(\rho^\alpha+2N)
(\rho^\alpha+2W)}$, -- remains small everywhere for small $g$ and
large $N$ and $W$, \i.e.\ for large $n$ and $w$. Thus in the leading
approximation we can ignore string loop corrections.
In order to analyze the effect of the $\alpha'$ corrections on the
near horizon geometry, we focus on the region:
\be \label{enear}
\rho^\alpha << N, W\, .
\ee
In this region
the solution \refb{e6+}, \refb{estring}  takes the form:
\ben \label{e8}
ds_{string}^2 &=& -{(D-3)^2 \, \Omega_{D-2}^2\,
\rho^{2\alpha} \over 64\pi^2  n w} \, dt^2 + g^{2\gamma}
\, d\vec{x}^2
\, ,
\nonumber \\
S &=& 8\pi g^{-2+\gamma} \, {1\over (D-3)\Omega_{D-2}}\, \rho^{-\alpha}
\, \sqrt{nw}\, , \nonumber \\
T &=& \sqrt{n\over w} \, , \nonumber \\
F^{(1)}_{\rho t}  &=& {(D-3) \Omega_{D-2}\over 16\pi n}\, \alpha
\, \rho^{\alpha-1} \, , \nonumber
\\
F^{(2)}_{\rho t}  &=& {(D-3) \Omega_{D-2}\over 16\pi w} \,
\alpha \, \rho^{\alpha-1}.
\een
We now introduce rescaled coordinates:
\be \label{e9}
\vec y = g^\gamma
\, \vec x, \qquad r = \sqrt{\vec y^2} = g^\gamma \, \rho\, , \qquad
\tau = g^{-\alpha\gamma} \, {(D-3)\Omega_{D-2}\over 4\pi}
t / \sqrt{nw}\, .
\ee
In this coordinate system \refb{e8} takes the form:
\ben \label{e10}
ds_{string}^2 &=& -{r^{2\alpha} \over 4} \, d\tau^2 + \, d\vec{y}^2
\, , \qquad r^2 \equiv \vec y^2\, ,
\nonumber \\
S &=& {8\pi \over (D-3) \Omega_{D-2}}\, {\sqrt{nw}\over r^{\alpha}}
\, , \nonumber \\
T &=& \sqrt{n\over w} \, , \nonumber \\
F^{(1)}_{r\tau} &=& {1\over 4} \, \alpha \, r^{\alpha-1}
\, \sqrt{w\over n}\, , \nonumber
\\
F^{(2)}_{r \tau} &=& {1\over 4} \, \alpha \, r^{\alpha-1}\,
\sqrt{n\over w} \, ,
\een
where we have used the fact that $\alpha\gamma-2+\gamma=0$ for $\gamma$
and $\alpha$ given in \refb{edefgmn} and \refb{e6-} respectively.
Notice that in this new coordinate system the solution near the horizon is
determined completely by the charge quantum numbers $n$ and $w$ and is
independent of the asymptotic values of the moduli $g$ and $R$. This is an
example of the
attractor mechanism for supersymmetric black
holes\cite{9508072,9602111,9602136}, generalized to higher dimensions.

We now note that the
tree level low energy effective action involving charge neutral fields is
invariant under a rescaling of the form:
\be \label{e11}
G^{(10)}_{99} \to e^{2\beta} G^{(10)}_{99}, \qquad G^{(10)}_{9\mu}\to
e^{\beta} G^{(10)}_{9\mu}, \qquad B^{(10)}_{9\mu}\to
e^{\beta} B^{(10)}_{9\mu}\, , \qquad \Phi^{(10)}\to \Phi^{(10)}+\beta\, .
\ee
Physically this corresponds to a rescaling of
the compactification radius by $e^\beta$ (together with a shift in the
dilaton to compensate for
the effect of the change of the volume of the compact
manifold).
Clearly the full string theory is
sensitive to the radius of compactification
and is not invariant under this
transformation. However the tree level effective action involving charge
neutral fields, which are involved in the construction of the black hole
solution, is not sensitive to the compactification radius, and the action
as well as all the quantities ({\it e.g.} the black hole entropy) computed
from the effective action will be unchanged under this rescaling. In terms
of the four dimensional fields defined in \refb{e3} this amounts
to:
 \be
\label{e12}
T\to e^\beta T, \qquad A^{(1)}_\mu \to e^{-\beta}  A^{(1)}_\mu, \qquad
A^{(2)}_\mu \to e^{\beta}  A^{(2)}_\mu \, , \qquad \Phi\to\Phi\, ,
\qquad G_{\mu\nu}\to G_{\mu\nu}\, .
\ee

Next we recall that the complete tree level effective action
of the
heterotic string theory in the subsector under study has the form:
\be \label{e15}
\SSS = \int d^D x \, \sqrt{-\det G} \, S \, \LL(G_{\mu\nu},
T, A_\mu^{(1)}, A_\mu^{(2)}, \p_\mu S / S, \cdots)\, ,
\ee
where $\cdots$ stand for derivatives of various fields which have appeared
explicitly in the argument of $\LL$.
Under multiplication of $S$ by a constant, the
action gets multiplied by the same constant. This shows that given any
solution of the full equations of motion derived from the action
\refb{e15}, we can get another solution
by the transformation:
\be \label{esscale}
S\to K \, S\,
\ee
for an arbitrary constant $K$, leaving the string metric and
other fields unchanged. Since
this transformation
multiplies the action by a constant factor $K$, it also multiplies
the entropy associated with the black hole by the same factor.

Choosing $e^\beta=\sqrt{w/n}$ in \refb{e12} and $K=1/\sqrt{nw}$
in \refb{esscale} we can map the solution \refb{e10}
to:\footnote{We would
like to emphasize that the hatted solution is
related to the original solution \refb{e8} by transformations
which are exact symmetries of the equations of motion of tree level string
theory, but are not exact symmetries of the full string theory.}
\renewcommand{\wh}{\hat}
\ben \label{e16}
\wh{ds}_{string}^2 &=& -{r^{2\alpha} \over 4} \, d\tau^2 + \, d\vec{y}^2
\, , \qquad r^2 = \vec y^2\, ,
\nonumber \\
\wh S &=& {8\pi \over (D-3) \Omega_{D-2}} \,
{1 \over r^\alpha}\, , \nonumber \\
\wh T &=& 1 \, , \nonumber \\
\wh F^{(1)}_{r\tau} &=& {1\over 4} \, \alpha \,
r^{\alpha-1} \, , \nonumber
\\
\wh F^{(2)}_{r \tau} &=& {1\over 4} \, \alpha \, r^{\alpha-1} \, .
\nonumber
\\
\een

We now note that
\begin{itemize}
\item The solution has no dependence
on
any parameter and is completely universal.
\item \refb{e16} is an exact solution of the classical low energy
supergravity equations of motion. This follows from the fact that \refb{e6+},
\refb{estring}
is a solution of these
equations for all $n$ and $w$, and \refb{e16}
is obtained from this solution by taking the limit $n,w\to\infty$ and
carrying out operations which are exact symmetries of the
classical low energy
supergravity equations of motion.
\item For $r>>1$ the higher derivative corrections to the solution
\refb{e16}
are small. This can be seen by introducing a new
coordinate $\eta$ via the relation
$\tau=2\eta / r^\alpha$, and writing the solution as
\ben \label{etrssol}
\wh {ds}_{string}^2 &=&   -d\eta^2 + \, d\vec{y}^2 +2\, \alpha\,
{\eta\over r} \,
d\eta dr
- \alpha^2\, {\eta^2\over r^{2}} dr^2
\, , \qquad r^2 \equiv \vec y^2\, , \qquad \alpha\equiv D-3\, ,
\nonumber \\
\p_r \wh S / \wh S&=&  -\alpha/r \, , \nonumber \\
\wh T &=& 1 \, , \nonumber \\
\wh F^{(1)}_{r\eta} &=& {\alpha \over 2\, r} \, , \nonumber
\\
\wh F^{(2)}_{r \eta} &=& {\alpha  \over 2\, r} \, .
\nonumber
\\
\een
{}From this we see that for fixed $\eta$, the metric
approaches flat metric and all
other fields become trivial for large $r$. Hence the corrections
due to higher derivative terms are small in this region
and we expect the solution of the complete classical
equations of
motion of string theory to be approximated
by \refb{e16} in this limit.

\end{itemize}

Since \refb{e16} has a completely
universal form without any parameter. and since furthermore the action
\refb{e15} is also completely universal, it is clear that the higher
derivative terms in \refb{e15} will change \refb{e16} to a universal form:
\ben \label{ei1}
\wh{ds}_{string}^2 &=& - e^{2G(\xi)} \, d\tau^2 + d\xi^2
+ e^{2 F(\xi)} \, d\Omega_{D-2}^2 \, ,
\nonumber \\
\wh S &=& e^{\sigma(\xi) -G(\xi) - (D-2) F(\xi)} \, , \nonumber \\
\wh T &=&  e^{\chi(\xi)} \, , \nonumber \\
\wh A^{(1)}_{\tau} &=& \psi_1(\xi) \, , \nonumber
\\
\wh A^{(2)}_{\tau} &=& \psi_2(\xi) \, ,
\een
where $d\Omega_{D-2}^2$ denotes the metric on the unit $(D-2)$-sphere
and $G$, $F$, $\sigma$, $\chi$, $\psi_1$ and $\psi_2$ are a set of
universal functions of the radial coordinate $\xi$. \refb{ei1}
represents a particular parametrization of
the most general spherically symmetric configuration; this
parametrization has been chosen for later
convenience. For large
$\xi$ \refb{ei1} must agree with the solution \refb{e16} for suitable choice
of relation between $r$ and $\xi$.
Using the inverse of the transformation \refb{esscale} and \refb{e12}
we can now generate the
modified version of the solution \refb{e10}:
\ben \label{e2.23a}
{ds}_{string}^2 &=&  - e^{2G(\xi)} \, d\tau^2 + d\xi^2
+ e^{2 F(\xi)} \, d\Omega_{D-2}^2 \, ,
\nonumber \\
S &=& \sqrt{nw}  \, e^{\sigma(\xi) -G(\xi) - (D-2) F(\xi)}
\nonumber \\
T &=&  \sqrt{n\over w} \, e^{\chi(\xi)} \, \nonumber \\
A^{(1)}_{\tau} &=& \sqrt{w\over n} \, \psi_1(\xi)  \, , \nonumber
\\
A^{(2)}_{\tau} &=& \sqrt{n\over w} \, \psi_2(\xi)  \, .
\een

We now turn to the computation of entropy associated with this
solution. In the presence of higher derivative corrections the entropy is
no longer proportional to the area of the event horizon; there are
additional corrections\cite{9307038,9312023,9403028,9502009}. These
corrections all have the property that if the action is multiplied by a
constant then the entropy associated with a given solution also gets
multiplied by the same
constant. Now suppose $a$ denotes the entropy associated with the solution
\refb{ei1}. Then since the solution \refb{ei1} and the action \refb{e15}
are both universal, $a$ must be a purely numerical coefficient. Since
the transformation \refb{esscale} with $K=1/\sqrt{nw}$ multiplies the action
by a factor of $1/\sqrt{nw}$, whereas the transformation \refb{e12} leaves
the action invariant, and since upon multiplying the action by a constant
factor the entropy associated with the solutions gets multiplied by a
constant factor, we see that
the entropy associated with the solution \refb{e2.23a} must be
given by:
\be \label{e20}
S_{BH} = a \sqrt{nw}\, .
\ee

On the other hand counting of states of fundamental heterotic string
carrying $w$ units of winding and $n$ units of momentum along $S^1$ shows
that for large $n$ and $w$ the degeneracy of states grows as $e^{4\pi
\sqrt{nw}}$\cite{rdabh0}.
Thus the statistical entropy, defined as the logarithm of the
degeneracy of states, is given by:
 \be \label{e21}
S_{stat} \simeq 4\pi\sqrt{nw} \, ,
\ee
for large $n$ and $w$. Thus we see that up to an overall
multiplicative constant the
statistical entropy agrees with the Bekenstein-Hawking entropy of the
black hole.

\sectiono{Modification of the Solution by Higher Derivative Terms and
its Near Horizon Limit} \label{scomplete}

In this section we shall propose an ansatz for how the solution
\refb{e6+} is modified by the higher derivative
corrections to the effective action close to the horizon of the
black hole. In order to simplify the formul\ae\ we shall focus on
the modification of the hatted solution \refb{e16}, since
once we find the modification of this
solution, we can determine the modification of the original solution
as in \refb{e2.23a}.

Since the solution in the supergravity limit
is invariant under the $(D-1)$-dimensional rotation group $O(D-1)$,
we can restrict our search for the modified solution within the class
of rotationally invariant configurations. If we now assume that there is
some choice of field variables (which need not be the variables which appear
directly in the $\sigma$-model describing string propagation in this
background)
in which the horizon acquires a finite size, then it must have the shape
of $S^{D-2}$ and we expect the near horizon geometry to be a product
of $S^{D-2}$ and a two dimensional space of Lorentzian signature. Symmetry
considerations do not tell us what this two dimensional space might be;
however for $D=4$ the analysis of
\cite{0409148,0410076,0411255,0411272,0501014,0502126,0502157,0504005}
taking into account a special
class of higher derivative terms shows that this two dimensional
space is $AdS_2$. We shall assume that even for $D>4$ the two
dimensional space is $AdS_2$. Since $AdS_2$ has symmetry group SO(2,1),
it is natural to postulate that not only the metric but also the other
near horizon field configurations are invariant under this SO(2,1)
symmetry. This fixes the form of all the near horizon
field configurations up to a
few constants:\footnote{\label{f1} Note that the `string metric' does not necessarily
refer to the metric that appears in the $\sigma$-model describing string
propagation in this background. Rather it simply implies that the metric
appearing in \refb{emod} has the same scaling properties as the string
metric under the transformations \refb{e12}, \refb{esscale}.}
\ben \label{emod}
\wh{ds}_{string}^2 &=& v_1\left(- r^2 \, d\tau^2 + {1\over r^2} dr^2 \right)
+  v_2 d\Omega_{D-2}^2\, ,
\nonumber \\
\wh S &=& v_3, \nonumber \\
\wh T &=& v_4 \, , \nonumber \\
\wh F^{(1)}_{r\tau} &=& v_5 \, , \nonumber
\\
\wh F^{(2)}_{r \tau} &=& v_6\, .
\nonumber
\\
\een
Here $v_1,\ldots v_6$ are
constants to be determined by solving the equations
of motion. For consistency we must check that there is a solution that
interpolates between \refb{emod} for small $r$ and \refb{e16} for
large $r$, but we shall not attempt to analyze this question here.
Instead we shall focus on the analysis of the equations of motion near
the horizon. Some aspects of the interpolating solution will be
discussed in section \ref{sinter}.

First we analyze the equations of motion of the gauge fields.
{}From eq.\refb{e15} and spherical symmetry of the configuration
it follows that these equations take the
form:\footnote{This form of the equations of motion requires
that the Lagrangian density depends only on the gauge field strengths
and not for example on the gauge fields themselves. Thus inclusion of
Chern-Simons forms in the action would modify this structure. Since
for the solution under study the anti-symmetric three form field strength
$H_{\mu\nu\rho}$ vanishes, we can use the form of the equation given in
\refb{ex1}.}
\be \label{ex1}
\p_r \left[\sqrt{-\det G} \, S \, {\p \LL\over \p F^{(i)}_{r\tau}}
\right] = 0\, , \qquad i=1,2\, .
\ee
Thus the combination $\left[\sqrt{-\det G} \, S \, {\p \LL\over \p
F^{(i)}_{r\tau}}
\right]$ is independent of $r$. Since for large $r$ the contribution
from higher derivative corrections to the action is small, we can
evaluate this expression at large $r$ using the supergravity action
\refb{e4+} and the form \refb{e16} for the
hatted solution, and then by \refb{ex1} the value
of this expression near the horizon must be the same. This gives us
two equations:
\be \label{ex2}
\sqrt{-\det \wh G} \, \wh S \, {\p \wh \LL\over \p
\wh F^{(i)}_{r\tau}} = {1\over 2 \Omega_{D-2}} \, \sqrt{\det h^{(D-2)}},
\qquad \hbox{for $i=1,2$}
\, ,
\ee
where $h^{(D-2)}_{ij}$ denotes the metric on a unit $(D-2)$-sphere, and
$\wh \LL$ denotes $\LL$ evaluated for the hatted solution.

Let us define the function $f(v_1,\ldots v_6)$ through the relation:
\be \label{ex3}
\left.
\sqrt{-\det \wh G} \, \wh S \, \LL(\wh G_{\mu\nu},
\wh T, \wh A^{1)}_\mu, \wh A^{(2)}_\mu, \p_\mu
\wh S/\wh S=0)\right|_{r\simeq 0}
= {1\over \Omega_{D-2}}\, \sqrt{\det h^{(D-2)}}\,  f(v_1,\ldots v_6)\, ,
\ee
where the hatted configuration is as given in \refb{emod}.
Then eqs.\refb{ex2} may be written as:
\be \label{ex4}
{\p f\over \p v_5} = {1\over 2}\, , \qquad  {\p f\over \p v_6} ={1\over 2}\, .
\ee

We now turn to the equations of motion for the metric and the scalar fields
$S$ and $T$. Due to the $SO(2,1)\times SO(D-1)$
symmetry of the configuration, the metric equation
reduces to two independent
scalar equations; obtained by extremizing the action with respect
to any one of the components of the metric along $AdS_2$, and any one of the
components of the metric along $S^{D-2}$.
The scalar field equations are obtained
by extremizing the action with respect to these fields. The resulting
set of equations may be written as\footnote{This is possible because
we have chosen the most general configuration consistent with the $SO(2,1)
\times SO(D-1)$
symmetry. Furthermore,
in
the coordinate system that we have chosen, the symmetry transformation
laws are independent of the parameters $\{v_i\}$.}
\be \label{ex5}
{\p f \over \p v_i}=0\, , \qquad 1\le i\le 4\, .
\ee
Eqs.\refb{ex4} and \refb{ex5} give complete set of equations which need to be
solved in order to compute the parameters $v_i$.

Given the solution, we can compute the entropy associated with the solution
by the formula given in \cite{9307038,9312023,9403028,9502009}.
In the present context this formula gives:
\be \label{ex6}
\wh S_{BH} = \left. 8\pi \, \wh S \, {\p \wh\LL\over \p \wh R_{Grtrt}}
\wh G_{rr} \wh G_{tt} \wh A_{D-2}\right|_{r=0}
\, ,
\ee
where $\wh R_{G\mu\nu\rho\sigma}$ denotes the
Riemann tensor computed using
the string metric $\wh G_{\mu\nu}$,\footnote{We could use any other metric,
{\it e.g.} the canonical metric for this computation as long as we use the
same metric everywhere in eq.\refb{ex6}. In computing the derivative in
\refb{ex6} we regard different components of the Riemann tensor as independent
variables.}
$\wh A_{D-2}$ is the area of the event horizon
measured in the string metric
$\wh G_{\mu\nu}$,
and the hat on top of $S_{BH}$ denotes that we are
computing the entropy associated with the hatted solution.
Thus $\wh S_{BH}$
can be identified with the constant $a$ appearing in \refb{e20}.
For the background \refb{emod}
\be \label{ex7}
\wh A_{D-2}=(v_2)^{(D-2)/2} \Omega_{D-2}, \qquad
\wh G_{rr}=v_1/r^2, \qquad
\wh G_{tt}=-v_1 r^2\, , \qquad \wh S=v_3\, ,
\ee
and we have
\be \label{ex8}
a = \wh S_{BH}=\left.
-8\pi \, (v_1)^2 \, (v_2)^{(D-2)/2} \, v_3\, \Omega_{D-2}
\,
{\p \wh \LL\over \p \wh R_{Grtrt}}\right|_{r=0} \, .
\ee
This expression can be simplified using eq.\refb{ex3}. If we define 
$f(\lambda;v_1,\cdots v_6)$ through a similar equation by multiplying
in the expression for $\LL$
every factor of $\wh R_{G\alpha\beta\gamma\delta}$ for 
$\alpha,\beta,\gamma,\delta=r,t$
by a factor of $\lambda$, then we have the relation:
\be \label{ex3a}
\left.
\sqrt{-\det \wh G} \, \wh S \, \wh R_{G\alpha\beta\gamma\delta}\,
{\p\wh\LL\over \p\wh R_{G\alpha\beta\gamma\delta}}
\right|_{r\simeq 0}
= \left. {1\over \Omega_{D-2}}\, \sqrt{\det h^{(D-2)}}\,  {\p f(\lambda;
v_1,\ldots v_6)\over \p\lambda}\right|_{\lambda=1}\, .
\ee
Using the relation:
\be \label{erel1}
R_{G\alpha\beta\gamma\delta} = - v_1^{-1}\, (G_{\alpha\gamma}G_{\beta\delta} 
- G_{\alpha\delta} G_{\beta\gamma})\, , \quad \alpha,
\beta,\gamma,\delta=r,t\, ,
\ee
and the expression for $G_{\mu\nu}$ given in eqs.\refb{emod}, we can rewrite
\refb{ex3a} as
\be \label{erel2}
\left. {\p\wh\LL\over \wh R_{Grtrt}}
\right|_{r\simeq 0} = {1\over 4}\, v_1^{-2}\, v_2^{-(D-2)/2} \,
v_3^{-1} \, {1\over \Omega_{D-2}}\, \left. {\p f(\lambda;
v_1,\ldots v_6)\over \p\lambda}
\right|_{\lambda=1}\, .
\ee
Eq.\refb{ex8} now gives
\be \label{erel3}
a=-2\pi\, \left. {\p f(\lambda;
v_1,\ldots v_6)\over \p\lambda} \right|_{\lambda=1}\, .
\ee

In order to proceed further we need to know the explicit form
of $\LL$.
This however is not known. What we plan to do next is to consider a
special higher derivative correction to the
action which is known to exist in the tree level heterotic string theory
and study its effect on the solution under consideration.
This is the Gauss-Bonnet term\cite{rzwiebach,9610237}:\footnote{Effect
of Gauss-Bonnet term on black hole entropy has been studied earlier
in different context\cite{9711053,0112045}.}
\be \label{ex9}
\Delta S = {C \over 16\pi} \, \int d^D x \, \sqrt{-\det G} \, S \,
\left[ R_{G\mu\nu\rho\sigma} R_G^{\mu\nu\rho\sigma} - 4 R_{G\mu\nu}
R_G^{\mu\nu}
+ R_G^2
\right]
\ee
where $C$ is a constant. For heterotic
string theory
\be \label{ex10}
C=1
\ee
but we shall work with arbitrary $C$ so that we can identify the effect of the
higher derivative term by examining the $C$ dependence of the final formul\ae.
Combining \refb{e4+}, \refb{ex9} and comparing this with \refb{e15}, we get
\ben \label{ey1}
\LL &=& {1\over 32\pi} \bigg[ R_G
+ S^{-2}\, G^{\mu\nu} \, \p_\mu S \p_\nu S -  T^{-2} \,
G^{\mu\nu} \, \p_\mu T \p_\nu T
\nonumber \\
&& - T^2 \,
G^{\mu\nu} \, G^{\mu'\nu'} \, F^{(1)}_{\mu\mu'}
F^{(1)}_{\nu\nu'} - T^{-2} \,
G^{\mu\nu} \, G^{\mu'\nu'} \, F^{(2)}_{\mu\mu'}
F^{(2)}_{\nu\nu'} \nonumber \\
&& + 2 \, C \, \left\{ R_{G\mu\nu\rho\sigma} R_G^{\mu\nu\rho\sigma}
- 4 R_{G\mu\nu} R_G^{\mu\nu}
+ R_G^2
\right\}
\bigg] + \ldots \, ,
\een
where $\dots$ denotes other terms which we are not including in our analysis.
These include terms related to \refb{ex9} by supersymmetry as well as other
higher derivative terms.

Substituting \refb{emod} into \refb{ey1}, and using the definition
of $f(\vec v)$ given in eq.\refb{ex3}, we get
\ben \label{ey2}
f(v_1, \ldots v_6) &=& {\Omega_{D-2}\over 32\pi} \, v_1 \, v_2^{(D-2)/2}
\, v_3 \, \left[ -{2\over v_1} +{(D-2)(D-3)\over v_2} +
{2 \, v_4^2 \, v_5^2\over v_1^2} +{2 \, v_6^2 \over v_4^2 \, v_1^2} \right.
\nonumber \\
&& \left. +{2 \, C\over v_2^2} \, (D-2)(D-3)(D-4)(D-5)
-{8C\over v_1 v_2} (D-2) (D-3)\right]\, . \nonumber \\
\een
The solutions to \refb{ex4}, \refb{ex5} are now given by
\ben \label{ey3}
v_2 &=& 4\, C \, [ (D-2)(D-3) - (D-4)(D-5)] \, , \nonumber \\
v_1 &=& {2 \, v_2 \over (D-2)(D-3)} \, , \nonumber \\
\wt v_3 \equiv {\Omega_{D-2}\over 32\pi} \, v_1 \, v_2^{(D-2)/2} \, v_3
&=& \left[ {16\over v_1^3 v_2} (v_2 + 4\, C \, (D-2) (D-3))\right]^{-1/2}\, ,
\nonumber \\
v_4 &=& 1\, , \nonumber \\
v_5 = v_6 &=& {v_1^2 \over 8 \wt v_3} \, .
\een
Finite values of $v_1$, $v_2$ and $v_3$
shows that the inclusion of the higher
derivative terms \refb{ex9} into the action does stretch the horizon of the
black hole.

Finally we turn to the computation of entropy using
\refb{erel3}. For the lagrangian density \refb{ey1},
$f(\lambda;v_1,\cdots  v_6)$ is given by\footnote{$f(\lambda,\vec v)$ can be
obtained from $f(\vec v)$ by the rescaling 
$v_1\to\lambda^{-1}v_1$, $v_2\to v_2$,
$\wt v_3\to \wt v_3$, $v_4\to v_4$, $v_5\to\lambda^{-1}v_5$ and $v_6\to
\lambda^{-1}v_6$.}
\ben \label{ey2a}
f(\lambda;v_1, \ldots v_6) &=& {\Omega_{D-2}\over 32\pi} \, v_1 \, v_2^{(D-2)/2}
\, v_3 \, \left[ -{2\lambda\over v_1} +{(D-2)(D-3)\over v_2} +
{2 \, v_4^2 \, v_5^2\over v_1^2} +{2 \, v_6^2 \over v_4^2 \, v_1^2} \right.
\nonumber \\
&& \left. +{2 \, C\over v_2^2} \, (D-2)(D-3)(D-4)(D-5)
-{8C\lambda\over v_1 v_2} (D-2) (D-3)\right]\, . \nonumber \\
\een
Eqs.\refb{erel3} and \refb{ey3} now give
\ben \label{ey5}
a &=& 4\pi {\wt v_3\over v_1\, v_2} \, \left( v_2 + 4\, C\, (D-2)\,
(D-3)\right)
\, , \nonumber \\
&=& \pi \, \sqrt C \, \sqrt{2\over (D-2)(D-3)} \, \{
8 (D-2)(D-3) - 4(D-4)(D-5)\}^{1/2}\, , \quad C=1\, . \nonumber \\
\een
This differs from the expected value $4\pi$ calculated using the statistical
entropy except for $D=4$ and $D=5$.\footnote{The agreement for $D=4$ is
probably related to the result of \cite{9711053} where a similar agreement
was found for large black holes carrying both electric and magnetic
charges.}
This is not a surprise since unlike in
\cite{0409148,0410076,0411255,0411272,0501014,0502126,0502157,0504005}
our analysis does not
include the complete set of terms needed to supersymmetrize the action;
-- the agreement with the correct answer for $D=4$ and $D=5$ is
most likely an accident.
It is
however encouraging that the effect of part of the higher derivative terms do
modify the geometry so as to yield a finite value of the entropy of these
black holes.

Before concluding this section we would like to mention some subtle
points in the analysis.
\begin{enumerate}
\item First of all note that the gauge field fluxes, encoded in the non-zero
constants appearing on the right hand side of eqs.\refb{ex4}, are crucial for
getting a non-trivial solution of the equations of motion. If the right hand
sides of eqs.\refb{ex4} had been zero, we would get $v_5=v_6=0$.
Eqs.\refb{ex5} with $f$ given in \refb{ey2} would
then imply that $v_1$ and $v_2$ must be infinite. This corresponds to the
flat space
limit.
\item By examining the structure of the solution \refb{emod} one
would be tempted to conclude that string propagation in this background is
described by a direct sum of two conformal field theories (CFT); -- one
associated with the $r,\tau$ coordinates labelling $AdS_2$ and the other
associated with the angular coordinates labelling $S^{D-2}$. The background
gauge fields only affect the CFT associated with the $r,\tau$ coordinate since
only the $r,\tau$ component of the gauge fields are non-zero. This would give
rise to
a coupling between the CFT associated with the $r,\tau$
coordinate and the CFT associated with the compact circle $S^1$. Had this
picture been correct, the equation determining the radius of $S^{D-2}$ would
be governed by the requirement of conformal invariance of the two
dimensional field theory involving the angular coordinates, and hence would
not involve the gauge fields. But we have just argued that if the gauge fields
vanish then there is no non-trivial solution for $v_2$, and so the equation
determining the radius of $S^{D-2}$ does
involve the gauge fields. The resolution of
this puzzle lies in the fact that the metric $G_{\mu\nu}$ appearing in the
solution \refb{emod} need not be the same metric that appears in the
$\sigma$-model describing string propagation in this background; instead two
metrics may be related by complicated field redefiniton. This point has already
been emphasized earlier in footnote \ref{f1}.

\end{enumerate}

\sectiono{On the Interpolating Solution} \label{sinter}

In this section we shall make a few remarks about the construction
of the full solution that
interpolates between the solution \refb{emod} near the horizon
and \refb{e16} for large $r$. The general form of the solution is
given in \refb{ei1}.
The large $r$ solution \refb{e16} corresponds to the choice
\be \label{ei2}
r = \xi, \quad e^G={r^\alpha\over 2}, \quad e^F=r, \quad e^{\sigma}
= {4\pi r^{D-2} \over (D-3) \Omega_{D-2}}, \quad \chi=0, \quad
\psi_1 = \psi_2 = {r^\alpha\over 4}+\hbox{constant}\, .
\ee
in eq.\refb{ei1}.
On the other hand the small $r$ solution \refb{emod}
corresponds to the
choice
\ben \label{ei3}
r= e^{\xi/\sqrt{v_1}}, \quad e^G=\sqrt{v_1}r, \quad e^F=\sqrt{v_2},
\quad e^\sigma = v_1^{1/2} v_2^{(D-2)/2} v_3\, r, \nonumber \\
\chi=\ln v_4, \quad \psi_1= v_5\, r+\hbox{constant}, \quad
\psi_2=v_6 \, r +\hbox{constant} \, .
 \een
 Our aim is to find a
solution that interpolates between \refb{ei2} for large $r$ and
\refb{ei3} for small $r$.\footnote{For a recent discussion on the
subtleties in constructing black hole solutions in the presence of
higher derivative terms, see \cite{0408200}.} For this we can
substitute the general form of the solution \refb{ei1} into the
action given in \refb{e15}, \refb{ey1}. After performing the
angular integration and some integration by parts in the $\xi$
variable this gives:\footnote{For studying $\alpha'$ corrections
to the D-brane solutions such reduction techniques have been used
earlier in \cite{0302136}.}
 \ben \label{ei4}
\SSS &=& \int dt \, d\xi \, \wt\LL\, , \nonumber \\
\wt\LL &=& {\Omega_{D-2}\over 32\pi} \, e^\sigma\left[
-(G')^2 - (D-2)\,
(F')^2 + (D-2)(D-3) e^{-2F} +(\sigma')^2 -(\chi')^2 \right .
\nonumber \\
&& +2 \, e^{2(\chi-G)} \, (\psi_1')^2 + 2 \, e^{-2(\chi+G)}
\,  (\psi_2')^2
\nonumber \\
&& + 8C(D-2)(D-3) \left\{ \left((F')^2 - e^{-2F}\right) \left(
(G')^2 +(D-2) F'G' -\sigma' G'\right) \right. \nonumber \\
&& -{1\over 3}(D-4) \sigma' (F')^3
+(D-4) \sigma' F' e^{-2F}
+(D-4) (F')^2 \left( (F')^2 -3\, e^{-2F}\right) \nonumber \\
&& \left. \left.  +{1\over 4} (D-4)(D-5)
\left( (F')^2 - e^{-2F}\right)^2 \right\} \right]  \, .
\een
Here $\prime$ denotes derivative with respect to $\xi$. Note that the
action does not contain any term involving two or more derivatives of
the fields. This is a consequence of the fact that we have chosen
the curvature squared terms in the Gauss-Bonnet form\cite{rzwiebach}.
The implication of this to the present problem will be discussed
shortly.

Besides having to satisfy the equations of motion derived
from the action \refb{ei4}, the fields $F$, $G$, $\sigma$, $\chi$, $\psi_1$
and $\psi_2$ must also satisfy the constraints associated with the
residual gauge symmetry $\xi\to\xi+\hbox{constant}$ in the parametrization used
in eq.\refb{ei1}. This amounts to setting the $\xi\xi$ component of the energy
momentum tensor, computed from the action \refb{ei4}, to zero. In terms of the
Lagrangian density $\wt\LL$ defined in \refb{ei4}, the constraint equation
takes the form:
\be \label{econs}
F'\, {\p\wt\LL\over \p F'} + G'\, {\p\wt\LL\over \p G'} +
\sigma'\, {\p\wt\LL\over \p \sigma'} + \chi'\, {\p\wt\LL\over \p \chi'}
+ \sum_{i=1}^2 \, \psi_i'\, {\p\wt\LL\over \p \psi_i'} - \wt \LL = 0\, .
\ee

It is easy to check that for the $v_i$'s given in \refb{ey3},
\refb{ei3} gives an exact solution to the equations
of motion derived from the action \refb{ei4} and the constraint equation
\refb{econs}.
On the other hand \refb{ei2} is an approximate
solution to these equations for large $\xi$ where the effect of the terms
proportional to $C$ is negligible.
We need to find a solution to the equations of motion for $F$, $G$, $\sigma$,
$\chi$, $\psi_1$ and $\psi_2$ derived from this action
that interpolates between \refb{ei2} for large $\xi$
and \refb{ei3} for small $\xi$. First we can try to carry out some consistency
checks by constructing
the Noether currents associated with various symmetries
of the problem and checking if the conservation laws are compatible
with our proposal for the near horizon geometry.
In particular, since the conservation law implies that the $\xi$ component
of the current should be $\xi$ independent, we can compare the $\xi$
components of the conserved currents for the configurations \refb{ei2} and
\refb{ei3} to see if they agree. Unfortunately this does not lead to
any non-trivial check. For example the conservation laws associated with the
shift symmetry of $\psi_1$ and $\psi_2$ are satisfied due to \refb{ex4}, while the
conservation law associated with the $\xi$ translation follows trivially from
the fact that the $\xi\xi$ component of the stress tensor vanishes at both ends
due to
\refb{econs}. Finally the conservation law associated with the
$G\to G+\lambda$, $\psi_i\to e^\lambda\psi_i$ symmetry can be satisfied by
adjusting the constant terms in the expression for the $\psi_i$'s
in eqs.\refb{ei2}, \refb{ei3}.

We now turn to the explicit analysis of the equations of motion and the
constraint equations. The
analysis may be simplified
by noting that the action \refb{ei4} has a symmetry under $\chi\to
-\chi$, $\psi_1\leftrightarrow\psi_2$. Since both the large and small $\xi$
solutions are invariant under this transformation, we can restrict
ourselves to the symmetric configuration:
\be \label{ei5}
\chi=0, \quad \psi_1=\psi_2\, .
\ee
The $\chi$ equation is now trivially satisfied.
$\psi_1$ and $\psi_2$ equations give:
\be \label{ei6}
e^{\sigma -2G} \, \psi_1' = e^{\sigma -2G} \, \psi_2' =
\hbox{constant} = {4\pi\over \Omega_{D-2}}\, ,
\ee
where the value of the constant has been fixed by studying the solution at
large $\xi$.

The non-trivial equations are those of $G$, $F$ and $\sigma$. Each of them
gives a second order differential equation by virtue of the fact that
the action contains only terms first order in the derivatives. Thus
the number of integration constants is the same as that in the absence
of higher derivative corrections, and
spurious oscillations seen in the interpolating solution at large
radius\cite{0411255,0411272} are absent here. Nevertheless we need six
integration constants (of which one can be eliminated due to the constraint
\refb{econs})
and it is not {\it a priori} clear that for a suitable choice of these
integration constants we get a
solution that interpolates between the two limiting forms \refb{ei2} and
\refb{ei3}.

In order to gain some insight into the problem we shall analyze
the linearized equations of motion around
the background \refb{ei2} for large
$r$ since a similar analysis turned
out to be instructive in case of four dimensional
string theories\cite{0411255}.
For this
we introduce fluctuations $g$, $f$ and $\wt\sigma$ around this background
through the equations:
\be \label{ess1}
e^G = {r^\alpha\over 2} \, e^g, \quad e^F = r\, e^f, \quad e^\sigma
= {4\pi r^{D-2}\over (D-3)\Omega_{D-2}} \, e^{\wt \sigma}, \quad \chi=0\, ,
\ee
with $\psi_1'$ and $\psi_2'$ given by eq.\refb{ei6}. The linearized
equations of motion
for the fluctuations $g$, $f$ and $\wt\sigma$, derived from the action
\refb{ei4}, then take the form:
\ben \label{ess2}
&&r^2 g'' + (\alpha+1) r g' - 2\alpha^2 g + \alpha r \wt\sigma' + 2\alpha^2
\wt\sigma=\OO(C)\, ,\nonumber \\
&&r^2 f'' + (\alpha+1) r f' + 2\alpha f + r \wt\sigma'=\OO(C) \, , \nonumber \\
&&\alpha r g' - \alpha^2 g + (\alpha +1) r f' + \alpha(\alpha+1) f
+ r^2 \wt\sigma'' + (\alpha+1) r \wt\sigma' + \alpha^2\wt\sigma = \OO(C)\, ,
\een
where $\OO(C)$ denotes the contribution from the Gauss-Bonnet
term in the action. On the other hand the constraint \refb{econs}
takes the form:
\be \label{ec1}
-\alpha\, r\, g'+\alpha^2 \, g - (\alpha+1)\, r\, f'
+\alpha\, (\alpha+1) \, f + (\alpha+1)\, r\, \wt\sigma' -\alpha^2\, \wt\sigma
=\OO(C)\, .
\ee
We shall first solve these equations ignoring these higher derivative terms
and then argue that their effect is small.
For $C=0$ the six linearly independent solutions of the equations
of motion \refb{ess2} are
found to be:
\ben \label{ess3}
&&g =1, \quad f=0, \quad \wt\sigma=1\, , \\
\label{ess4}
&&g=\alpha \,
r^{-1}, \quad f=r^{-1}, \quad \wt\sigma= (\alpha+1) \, r^{-1}
\, ,  \\ \label{ess5}
&&g=r^\alpha, \quad f=0, \quad \wt\sigma=0\, , \\ \label{ess6}
&&g=r^{-\alpha}, \quad f=r^{-\alpha}, \quad \wt\sigma=2\, r^{-\alpha}
\, , \\ \label{ess8}
&&g={3\alpha-1\over 3\alpha} \, r^{-2\alpha},
\quad f = {1\over (\alpha+1)}\, r^{-2\alpha}, \quad
\wt\sigma = r^{-2\alpha}\, ,
\\ \label{ess7}
&&g={\alpha\over 2\alpha-1}\, r^{-\alpha+1}, \quad
f = {\alpha-1\over \alpha+1} \, r^{-\alpha+1}, \quad \wt\sigma
=r^{-\alpha+1}\, . \een
The constraint equation \refb{ec1} eliminates
the solution \refb{ess7}.
{}Since for each solution
the fields $g$, $f$ and $\wt\sigma$
have a power law behaviour $r^\beta$ for
some $\beta$, the higher derivatives
terms involving these fields are suppressed by
powers of $1/r$ for large $r$. This justifies neglecting the correction
terms proportional to $C$ in this region.

A generic solution of the equations of
motion and the constraint equation
near the background \refb{ei2} will be given by an arbitrary linear
combination of the five solutions given in
\refb{ess3}-\refb{ess8}.
Some of these fluctuations can be recognized as familiar objects.
For example \refb{ess3} corresponds to a scaling of the time coordinate
$\tau$ and \refb{ess4} corresponds to a shift in the radial variable $r$.
Thus these modes are gauge artifacts.
\refb{ess5} represents the result of keeping correction terms of order
$\rho^\alpha$ in going from \refb{e6+}, \refb{estring} to \refb{e8}
(with $n=w$, $R=4$).
Since the original solution \refb{e6+}, \refb{estring} preserves half
of the space-time supersymmetry, it follows that the deformation
\refb{ess5} preserves half of the space-time supersymmetry. The other
two modes \refb{ess6}, \refb{ess8} are not familiar objects;
however note that each of these modes decays for large $r$. Also we expect
that these deformations do not preserve any space-time supersymmetry.

We now propose the following scenario for constructing the interpolating
solution. We deform the near horizon geometry \refb{ei3} by appropriate set
of
perturbations in the fields $G$, $F$ and $\sigma$ which
are non-singular as
$r\to 0$ but grows as $r$ increases. As we evolve the solution towards
large $r$, we expect that the solution will approach \refb{ei2} plus
some appropriate deformations described in \refb{ess3}-\refb{ess8}.
Of these \refb{ess3} and \refb{ess4} may be removed by scaling of $\tau$
and shift of $r$, whereas \refb{ess6} and
\refb{ess8} vanish for large $r$.
As a result
we only need to worry about the deformation \refb{ess5}. As already
pointed out, this is a physical deformation of the background
and should be present
in the full solution.\footnote{This
deformation helps us get out of the $1/r^\alpha$ dependence of the field $S$
as in \refb{e10}
and makes it approach a constant value for large $r$ as in \refb{e6+}.}
Thus in order that our ansatz for the near horizon geometry is consistent,
the non-singular deformations
of the background \refb{ei3} should
contain a set of parameters such that by
adjusting these parameters we can adjust the coefficient of the
deformation proportional to \refb{ess5} for large $r$.
In particular for a suitable
choice of parameters this coefficient may be made to vanish, and we should
recover the configuration \refb{ei2} for large $r$.

In principle we could analyze the non-singular deformations
of the background
\refb{ei3} for small $r$ and determine if there are enough parameters which
allow us to adjust the coefficient of \refb{ess5}. However since
the curvature and the other field strengths in \refb{ei3}
are of the order of the string
scale, this analysis is sensitive to the precise form of the higher derivative
terms, and could change when we add to the action other terms needed to
satisfy the requirement of space-time supersymmetry. For this reason we shall
not carry out the explicit analysis of this problem here.
We would like to note however that had
we started with the fully supersymmetric action, we could have tried
to find the interpolating solution by imposing the half-BPS condition.
Typically this would simplify the analysis since the BPS conditions give
first order equations for various fields instead of second order equations.
The construction of the full solution
would involve deforming the near horizon geometry \refb{ei3} by a
half-BPS perturbation that is non-singular
as $r\to 0$ but grows at large $r$, and
showing that for an appropriate choice of perturation parameters
the geometry approaches that given in \refb{ei2} for large $r$.
In fact in this case the deformations given in \refb{ess6}-\refb{ess8}
will be absent from the very beginning as they do not satisfy the
requirement of preserving half of the space-time supersymmetry.

\medskip

{\bf Acknowledgement}: I wish to thank A.~Dabholkar and N.~Iizuka
for discussion during the early stages of this work. I would also
like to thank the Abdus Salam International Centre for Theoretical
Physics and the Instituut voor Theoretische Fysica at the
University of Leuven for hospitality during the course of this
work.

\end{document}